\pdfoutput=1
\documentclass[sigconf,10pt]{acmart}

\copyrightyear{2026}
\acmYear{2026}
\setcopyright{cc}
\setcctype{by}
\acmConference[RE-NET '26]{First International Workshop on a Resilient, Renewable, and Responsible Internet}{December 07--10, 2026}{Utrecht, Netherlands}
\acmBooktitle{First International Workshop on a Resilient, Renewable, and Responsible Internet (RE-NET '26), December 07--10, 2026, Utrecht, Netherlands}
\acmDOI{10.1145/3843201.3856200}
\acmISBN{979-8-4007-2981-2/2026/12}

\usepackage{booktabs}
\usepackage{subcaption}
\usepackage{multirow}
\usepackage[inline]{enumitem}
\usepackage{xurl}

\usepackage{siunitx}
\usepackage{xspace}

\newcommand{\pkk}[1]{\num[round-mode=figures,round-precision=3]{\fpeval{(#1)/1000}}\,k}

\newcommand{\firstly}[0]{\textit{i)}\xspace}
\newcommand{\secondly}[0]{\textit{ii)}\xspace}

\makeatletter
\newcommand{\additionalaffilnote}{{%
  \renewcommand\thefootnote{\@fnsymbol\c@footnote}%
  \let\@footnotetext\@footnotetext@nolink
  \footnotetext[1]{Also with University of Twente.}}}
\makeatother

\begin{document}

\title[Split Skies]{Split Skies: Operational Experiences Extending a Multi-Cloud Network
	Telescope into the Asia-Pacific}

\author{Nils Kempen}
\affiliation{%
	\institution{University of M\"unster}
	\city{M\"unster}
	\country{Germany}
}
\email{}

\author{Benjamin Schumann}
\affiliation{%
	\institution{University of Twente}
	\city{Enschede}
	\country{The Netherlands}
}
\email{}

\author{Ralph Holz}
\affiliation{%
	\institution{University of M\"unster}
	\city{M\"unster}
	\country{Germany}
}
\additionalaffiliation{%
	\institution{University of Twente}
	\city{Enschede}
	\country{The Netherlands}
}
\email{}

\begin{abstract}

	Comprehensive visibility into Internet Background Radiation (IBR) requires
	diverse measurement infrastructure. Recent work showed that cloud IBR
	depends on the provider making single-provider telescopes systematically
	biased. These studies relied entirely on Western providers and left the
	Asia-Pacific region as a blind spot. In this report we extend a multi-cloud
	network telescope onto Tencent Cloud inside mainland China. We detail how
	Internet fragmentation, geo-politics, and provider rules complicate global
	measurement through strict onboarding barriers and an unreliable management
	plane. Once deployed, our nodes captured high-volume regional scanning
	invisible to our Western vantage points. We conclude by connecting our
	findings to the Responsible Internet framework. This framework relies on
	independent global observation to build a transparent Network Inspection
	Plane. We show that building this plane is bottlenecked by sovereign
	ecosystems and requires a collaborative community effort.

\end{abstract}

\ccsdesc[300]{Networks~Network security}
\ccsdesc[500]{Networks~Network measurement}

\keywords{Internet Background Radiation, Network Telescope, Cloud Computing,
	Internet Measurement, Internet Fragmentation}

\makeatletter\let\@authornotes\@empty\makeatother
\maketitle

\section{Introduction}
\label{sec:introduction}

A resilient Internet depends on our ability to observe it.
\additionalaffilnote%
The work by Hesselman et al. established that independent measurement is
essential to build a ``Responsible Internet'' (RI) and achieve network
transparency \cite{hesselmanResponsibleInternetIncrease2020}. Network telescopes
capturing Internet Background Radiation (IBR) (unsolicited traffic that reaches
any routed address) have long been used for observing events on the Internet.
They have allowed the detection of scanning campaigns, backscatter from
denial-of-service attacks, misconfigurations, as well as Internet outages
\cite{mannelLessonsLearnedOperating2025a} and
measuring these events provides the independent measurement data necessary to
support the transparency goals defined in the RI framework.

For two decades, \emph{classical network telescopes}~(darknets) that passively
capture traffic to large blocks of unused address space have been used for these
observations~\cite{mooreNetworkTelescopesTechnical2004}.
Unfortunately their sustainability is increasingly threatened. Operators reclaim
dark address blocks to meet demand for scarce IPv4 space. The largest publicly
available network telescope (UCSD-NT) run by CAIDA, for example, has shrunk to
\num{63.56}\% of a full /8~\cite{mannelLessonsLearnedOperating2025a}. At the
same time, scanners are aware of network telescope ranges and shift their scans
away from uniformly random targets toward dense, high-value cloud
infrastructure~\cite{izhikevichCloudWatchingUnderstanding2023,
richterScanningScannersSensing2019, pauleyDScopeCloudNativeInternet2023}.
In response, researchers have begun to examine the cloud itself as a vantage
point for IBR observation~\cite{pauleyDScopeCloudNativeInternet2023,
bortoluzziCloudTelescopeDistributed2023,
izhikevichCloudWatchingUnderstanding2023}. In our own prior work, we deployed
\textit{CLOUD-NT}, a passive multi-cloud network telescope spanning five
providers, and found that IBR in the cloud is \emph{provider-dependent
rather than geography-dependent}~\cite{kempenMultiCloudViewIBR2026}.
A telescope hosted on a single provider therefore carries a systematic
observational bias. This finding has a practical implication. If the provider
determines what a telescope sees, then the set of providers a researcher can
actually deploy on determines the limits of visibility. That set is shaped not
by science but by geo-politics, provider rules, and verification and payment
barriers.
These Internet fragmentation problems, already identified in
\cite{hesselmanResponsibleInternetIncrease2020}, restrict who can observe the
network and hinder diverse global observation needed to achieve infrastructure
transparency.
CLOUD-NT, like most prior cloud telescope work, was built entirely on Western
providers (AWS, GCP, Azure, Vultr, and DigitalOcean). For cloud-hosted
telescopes the Asia-Pacific region, and mainland China in particular, remained a
blind spot. To our knowledge, no prior cloud telescope deployment observed IBR
from within mainland China. We set out to extend our cloud-based network
telescope to additional providers in these regions.

This paper is an experience report on a second deployment wave of the CLOUD-NT
effort that set out to close that gap by extending the telescope eastward onto
Tencent Cloud. We describe what this took operationally and what the new vantage
point revealed. Our contributions are
\begin{enumerate*}[label=(\roman*)]
    \item an account of deploying passive telescope nodes in a
    politically constrained cloud ecosystem, covering a failed Alibaba Cloud
    attempt, the pivot to Tencent, per-provider infrastructure divergence, and
    in-country network interference, distilled into three general lessons
    (\textbf{L1} to \textbf{L3}) about Internet
    fragmentation~\cite{drakeInternetFragmentation2016}
    (\S\ref{sec:experience}),
    \item evidence that the Asia-Pacific vantage point captures high-volume
    activity that our earlier Western-provider deployments missed
    (\S\ref{sec:vantage}), and
    \item recommendations for building resilient, representative, and
        responsibly operated community monitoring infrastructure, which serves
        the RI as an independent observer.
\end{enumerate*}

\section{Background on CLOUD-NT}
\label{sec:background}

This report builds directly on our CLOUD-NT effort, and we recap only what is
needed to follow it, referring to~\cite{kempenMultiCloudViewIBR2026} for the
full design and analysis.
CLOUD-NT is a strictly
passive multi-cloud network telescope whose Terraform scripts provision the
smallest available virtual machine (VM) in as many regions and availability
zones as possible across five providers (AWS, GCP, Azure, Vultr, and
DigitalOcean), each listening on a single public IPv4 address. The security
groups forward all IPv4 ingress while host firewall rules silently drop every
reply, and management runs over IPv6. The first deployment ran \num{336} VMs
across \num{143} regions for two weeks in April 2025 and was compared against
two classical telescopes, the widely known UCSD-NT (a /9 and a /10) and a
publicly unknown one, SURF-NT (a /16)~\cite{kempenMultiCloudViewIBR2026}.

Two of its findings motivate this work.
\firstly IBR in the cloud is
\emph{provider-dependent rather than geography-dependent}. Traffic collected
within a single provider is more self-similar than traffic collected across
providers in the same region, so a single-provider telescope is systematically
biased. \secondly the scanner populations seen by cloud and classical telescopes
barely overlap. Only \num{0.6}\% of the \num{941}k scanners observed were common
to all three telescopes, and cloud IPs are disproportionately targeted on
cloud-service ports. Neither vantage point subsumes the other. The CLOUD-NT
discussion argued, on this basis, for multi-provider deployments and a
community-run cloud telescope with a data-governance framework. Yet the
providers it used were all Western. Whether provider-dependence extends into the
Asia-Pacific, and how far region matters there, was left open. This is the gap
the deployment described here set out to probe.

\section{Extending CLOUD-NT Eastwards}
\label{sec:experience}

Our goal was to extend CLOUD-NT's coverage into the Asia-Pacific, and mainland
China specifically, so that we could test whether its provider-dependence
finding held beyond the Western cloud market. The second deployment run
encompassed \num{28} VMs in total, \num{14} on Tencent Cloud (China and
Southeast Asia) and \num{14} on DigitalOcean (North America, Europe, and parts
of the Asia-Pacific), across \num{11} countries from May 18 to June 6, 2025
(\num{20} days). This deployment ran on these two providers only. DigitalOcean
served as the multi-region comparison provider that we already knew how to
operate, its VMs spanning North America, Europe, and parts of the
Asia-Pacific, while Tencent
Cloud served as the new Asia-Pacific vantage point. The
effort, and the lessons, were almost entirely on the Tencent side. We recount
them here because each is a concrete instance of a general obstacle that
Internet fragmentation and platform opacity place in the path of global
measurement. We number the lessons \textbf{L1} to \textbf{L3} and return to
them in \S\ref{sec:lessons}.

\paragraph*{\textbf{L1: Who gets to measure is constrained by locally determined
business practices}} 
Initially we chose to use Alibaba Cloud as our next cloud provider, because it
is the largest provider in the region and could give us a good view into Chinese
networks, but we never managed to deploy our network telescope there.
Provisioning infrastructure in mainland China required verification procedures
we could not easily satisfy from abroad. Even after we managed to successfully
create an account on Alibaba Cloud, it was quickly flagged for suspicious
activity
following our initial API calls.
It could only be unblocked with bank statements that the provider judged to
show insufficient funds.
Provisioning was severely delayed by documentation and management
consoles designed exclusively for a domestic market. Translation tooling helped
but could not eliminate the substantial operational friction.
In the end, we decided to abandon
Alibaba and pivoted to Tencent, the region's second-largest provider, where an
account could be established and the deployment made to work. Hence, the
technical design of the cloud telescope was not the bottleneck, but access to
the provider was.
For measurement research, this means the vantage points a
scientist may occupy are decided by provider policies across different
jurisdictions and strict identity requirements.
Such onboarding barriers create critical blind spots in global IBR analysis.

\paragraph*{\textbf{L2: The software is portable, the provisioning code is not}}
The software that runs on the VMs after provisioning, our capture and management
scripts, is fully provider-agnostic and moved onto Tencent relatively unchanged.
The Terraform provisioning code on the other hand had to be adjusted
per-provider, and even though CLOUD-NT already had support for five providers,
extending it to Tencent required substantial new code for a different resource
model, distinct API and security-group conventions, and an unfamiliar account
and billing structure. The general lesson is that provisioning code encodes the
inner workings of the ecosystems it was written for, so there is a larger
porting cost for distinctly new and different ecosystems.
For an RI, this implies that global observation infrastructure cannot be
developed unilaterally. Achieving global Internet transparency will require
active collaboration across regions to bridge these different technical
ecosystems.

\paragraph*{\textbf{L3: The management plane inside China is unreliable}} Even
once the VMs were running, operating them from outside was unreliable and we
faced problems that our other deployments did not have. Public IPv6, on which
CLOUD-NT relies for management and data-transfer, was inconsistently available
and required manual intervention on a per-VM basis. In-country network
conditions interfered with routine tooling. DNS resolution was blocked for some
destinations and access to \textit{GitHub} (which hosted some of our binaries)
was throttled, so provisioning steps that relied on this had to be reworked
around unreliable egress. These problems likely result from Chinese national
network controls. Previous research shows these controls are decentralized and
highly variable \cite{ensafiAnalyzingGreatFirewall2015}. This unpredictability
could explain our inconsistent network access and highlights why establishing
reliable observation nodes is so difficult.
During this manual setup we also misconfigured the VMs' firewall, allowing only
inbound TCP and UDP, blocking external ICMP on the Tencent side, whereas
DigitalOcean carried no such rule and received it. We flag the resulting ICMP
gap as a limitation of the dataset in \S\ref{sec:vantage}.
The same national filtering that disrupted our management scripts also
inherently filters the incoming background radiation we aim to capture.
Operating a telescope in these networks requires accounting for local
restrictions and data biases.

\section{What the New Vantage Point Sees}
\label{sec:vantage}

In this section we analyze what the new vantage point brings us in terms of
visibility and how it compares to one of our Western deployments.

\paragraph*{Method} We processed the raw packet captures into CAIDA
FlowTupleV3\footnote{\url{https://www.caida.org/projects/network_telescope/docs/data/flowtuple/}}
records with \textit{Corsaro}\footnote{\url{https://github.com/CAIDA/corsaro3}}
and report packet counts throughout, summed over the observation period from May
18 to June 6, 2025, which is \num{20} full UTC days. We approximate the three
IBR traffic classes from TCP flags. FlowTuple records carry per-packet header
information, so we assign a class to each packet on its own. A pure \textit{TCP
SYN} or any \textit{UDP} packet counts as scanning, a \textit{TCP ACK},
\textit{TCP RST}, or \textit{TCP FIN} as backscatter, and everything else,
including ICMP, as other.
On top of this per-packet classification we ran CLOUD-NT's sliding-window scan
detection, which labels per-source \num{15}-minute windows as port, host, or
random scan events~\cite{kempenMultiCloudViewIBR2026}. We determined scanner
detection thresholds by running the detection algorithm with different
thresholds, deciding on the one that lies on the elbow point of the detected
scanner counts, as we describe in detail in~\cite{kempenMultiCloudViewIBR2026}.
We validated this calibration against two blocklists, FireHOL's attacks
category\footnote{\url{http://iplists.firehol.org}} and the ISC SANS
research-scanner list\footnote{\url{https://isc.sans.edu/index_dyn.html}}. We
run the scanner detection
with a common threshold,
so that differences between the providers cannot stem from detection settings.
We exclude provider-internal traffic from special-use source addresses from
every result below. We further exclude two single-node, single-port streams
whose volume and shape match the reused-address spillover we described
in~\cite{kempenMultiCloudViewIBR2026}, because their massive volume would
otherwise distort the overall baseline.

\begin{figure}
	\centering
	\includegraphics[width=.8\columnwidth]{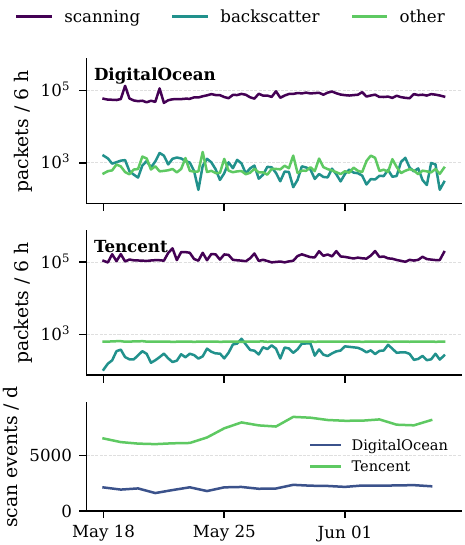}
	\caption{Packets per 6-hour bin by traffic class for each provider (top two
		panels, shared log scale) and detected scan events per day and provider
		(bottom panel).}
	\Description{Packets per 6-hour bin by traffic class for each provider (top
		two panels, shared log scale) and detected scan events per day and
		provider (bottom panel).}
	\label{fig:overview}
\end{figure}

\paragraph*{More pressure from a similar source pool} Averaged over the window,
the \num{14} DigitalOcean VMs together received \pkk{288759} $\pm$ \pkk{36079}
packets per day, of which \pkk{282941} $\pm$ \pkk{36978} (\num{98.0}\%) were
scanning. The \num{14} Tencent VMs received \pkk{539787} $\pm$ \pkk{80610}
packets per day, \pkk{535901} (\num{99.3}\%) of them scanning. Scanning traffic
accounts for almost this entire volume gap. The Tencent VMs saw \num{1.9} times
the packet volume of DigitalOcean, yet from a source pool of similar size.
DigitalOcean VMs observed \pkk{17252} $\pm$ \num{1173} distinct source IPs per
day against Tencent's \pkk{21577} $\pm$ \num{4927}, so Tencent drew \num{1.9}
times the packets from only about \num{1.3} times the sources. Sources send on
average more packets toward our Tencent VMs than they do to our DigitalOcean
VMs, \num{25.6} against \num{16.8} packets per source. The scan detection shows
a similar picture. We find \pkk{147581} scan events on the Tencent side against
\pkk{42223} on DigitalOcean, a factor of \num{3.5}. Figure~\ref{fig:overview}
shows that scanning dominates the traffic for both providers, that the Tencent
volume is steadier over time, and that the event asymmetry holds across the
whole window.

\begin{figure}
	\centering
	\includegraphics[width=.9\columnwidth]{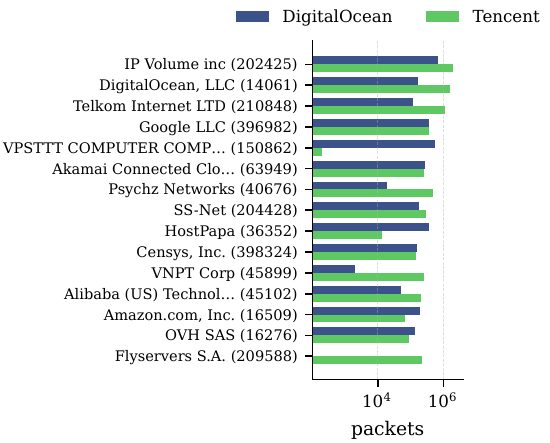}
	\caption{Top 15 source ASes by packet count, split by provider, log scale.}
	\Description{Top 15 source ASes by packet count, split by provider, log
		scale.}
	\label{fig:as}
\end{figure}

\paragraph*{Sources concentrate by provider} Among the top \num{30} source ASes
by packet count (Fig.~\ref{fig:as} shows the top 15), \num{12} of them, and
\num{5} of the top \num{10}, delivered more than \num{90}\% of their packets to
a single provider. The few ASes that split evenly are globally deployed
infrastructure networks such as Google (AS~396982) at \num{49.2}\% and Akamai
(AS~63949) at \num{47.8}\% Tencent share. This concentration is consistent with
CLOUD-NT's provider-dependence~\cite{kempenMultiCloudViewIBR2026} extending into
the Asia-Pacific, though in this setup provider and region coincide, so we
cannot yet tell whether sources target China as a region or Tencent as a
provider.
\begin{figure} \centering
    \includegraphics[width=.65\columnwidth]{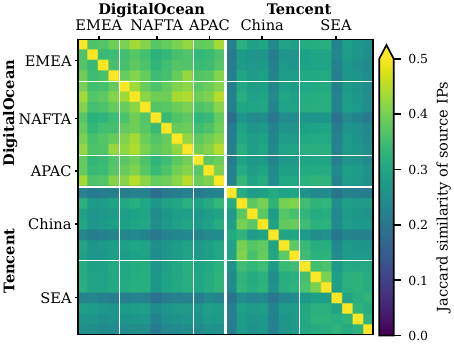}
    \caption{Pairwise Jaccard similarity of distinct source IP sets across 28
    VMs. Axes are grouped by provider and region. DigitalOcean spans EMEA,
    NAFTA, and APAC. Tencent spans China and SEA.}
    \Description{Pairwise Jaccard similarity of distinct source IP sets across
    28 VMs. Axes are grouped by provider and region. DigitalOcean spans EMEA,
    NAFTA, and APAC. Tencent spans China and SEA.} \label{fig:jaccard}
\end{figure}
The same pattern holds looking at individual source IPs.
Figure~\ref{fig:jaccard} shows the pairwise Jaccard similarity of the per-VM
source sets, mirroring an analysis from~\cite{kempenMultiCloudViewIBR2026}. Mean
pairwise similarity is \num{0.38} between DigitalOcean VMs, \num{0.30} between
Tencent VMs, and \num{0.27} across providers, with off-diagonal values from
\num{0.18} to \num{0.45}. The DigitalOcean VMs form a clear block, more similar
to one another (\num{0.38}) than to Tencent VMs (\num{0.27}). The Tencent VMs
are only weakly more similar to each other (\num{0.30}) than across providers,
so the block structure is asymmetric. Still, which sources a VM shares with
another is influenced more by provider than by location, extending the AS-level
concentration down to individual sources.

\begin{figure}
	\centering
	\includegraphics[width=.75\columnwidth]{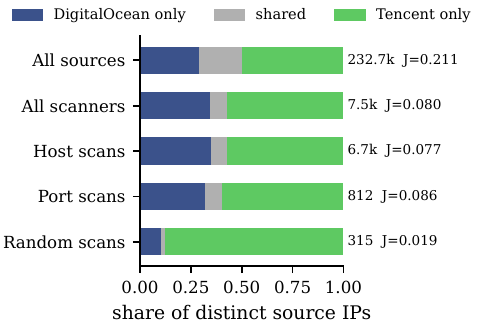}
	\caption{Overlap of distinct source IPs between deployments. Rows split the
		IPs into those unique to DigitalOcean, unique to Tencent, and shared. Values
		show the union size and Jaccard similarity for all sources, all scanners, and
		three scan types.}
	\Description{Overlap of distinct source IPs between deployments. Rows split
		the IPs into those unique to DigitalOcean, unique to Tencent, and shared.
		Values show the union size and Jaccard similarity for all sources, all
		scanners, and three scan types.}
	\label{fig:scanneroverlap}
\end{figure}

\paragraph*{The deployments detect largely disjoint scanner sets} The detected
scanner sets overlap only marginally. The scan detection identifies \num{3181}
scanner source IPs at DigitalOcean and \num{4896} at Tencent, of which only
\num{597} are shared, a Jaccard similarity of \num{0.08}
(Figure~\ref{fig:scanneroverlap}) against \num{0.21} across all public sources,
so the scanners overlap less than the overall IBR traffic. The two blocklists
used also cover the two populations unevenly. FireHOL's attacks category lists
\num{57.4}\% of the Tencent scanners but only \num{24.6}\% at DigitalOcean,
while the ISC SANS research-scanner list shows the opposite, \num{6.6}\% at
DigitalOcean against \num{2.5}\% at Tencent. Both lists together still miss most
DigitalOcean scanners (\num{71.1}\%) and a large share of Tencent's
(\num{41.3}\%).

\begin{figure*}
	\centering
	\includegraphics[width=.7\textwidth]{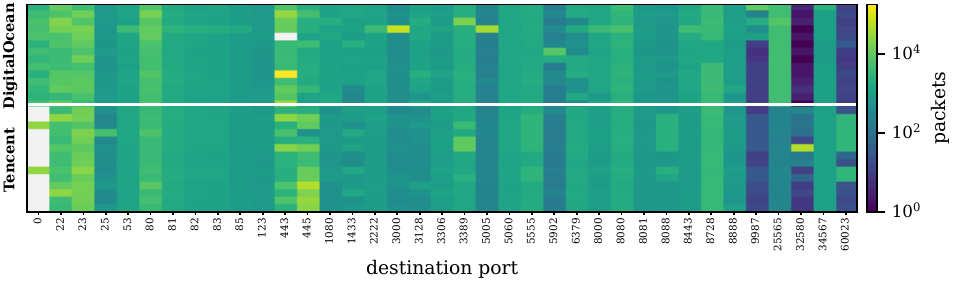}
	\caption{Packets per destination IP (rows, DigitalOcean block on top, Tencent
		below) for the union of each provider's \num{30} most-targeted destination
		ports (\num{37} total), log color scale, empty cells mean zero packets. The
		empty port-443 cell in the \texttt{lon1} row is an excluded SYN-flood
		stream.}
	\Description{Packets per destination IP (rows, DigitalOcean block on top,
		Tencent below) for the union of each provider's 30 most-targeted
		destination ports (37 total), log color scale, empty cells mean zero
		packets. The empty port-443 cell in the lon1 row is an excluded SYN-flood
		stream.}
	\label{fig:ports}
\end{figure*}

\paragraph*{Port traffic volumes vary by vantage point} While a network
telescope eventually receives IBR on all ports, the number of packets arriving
per port also depends on the vantage point (Figure~\ref{fig:ports}). Widely
scanned services reach both providers at similar levels, for example on ports
TCP/22 (SSH) and TCP/3389 (RDP) we observe roughly the same number of packets on
both providers ($\approx$\num{100}k and $\approx$\num{40}k respectively). Other
ports are different across providers. SMB port TCP/445 is heavily targeted on
Tencent, \pkk{179996} against \pkk{29551} packets, and Telnet port TCP/23 is
also more targeted on Tencent.
A few ports are single-node spikes rather than provider effects, such as
port~3000 with \pkk{64056} of its \pkk{75164} packets on one DigitalOcean node
(\texttt{fra1}) and port~32580 with \pkk{45104} of \pkk{46133} packets on one
Tencent node (\texttt{ap-guangzhou-6}).

\paragraph*{Region adds a second axis} While providers remain the dominant
factor influencing received IBR, geographic region adds a secondary axis of
influence. We observe a regional effect within individual providers. Our
DigitalOcean VMs in EMEA and NAFTA, for example, have very dissimilar scanner
populations. With only three VMs per region, though, we cannot draw any specific
conclusions from this observation. Clearer evidence is an event aimed at one region.
Between May 22 and 25 we observed a distributed UDP campaign from Psychz
Networks (AS~40676), and it targeted mainland China specifically. Over the full
observation window this AS sent \pkk{502451} packets to almost all UDP ports,
\num{89.2}\% of them in those four days. \pkk{480994} packets (\num{95.7}\%)
landed on the seven Tencent VMs inside mainland China and only \num{2590} on the
seven Tencent VMs outside it in Jakarta, Singapore, and Bangkok. Both sets of
VMs ran on Tencent, so this split is regional and not a provider effect.
DigitalOcean saw \pkk{18867} packets in total, its three Asia-Pacific VMs each
$\approx$\num{250}, no more than any other DigitalOcean VM. During this burst
\pkk{447220} packets landed on Tencent against \num{727} on DigitalOcean.
Neither provider diversity nor regional presence alone would have caught it.
Only a VM inside the targeted region did. This is a nation-targeted instance of
localized scanning, which Richter and Berger showed makes up a substantial share
of scan traffic and stays invisible to telescopes outside the targeted
region~\cite{richterScanningScannersSensing2019}.

\paragraph*{Traffic the platform injects} We also observed a large amount of
internal ICMP traffic on the Tencent VMs, sent by the platform itself. Every
Tencent VM received ICMP echo requests from the provider's link-local range
at a fixed rate of about one every three seconds, steady across the whole
observation period (\num{16.03} million packets in total from \num{14}
addresses). DigitalOcean saw no comparable packets. This is the signature of a
provider health check probing node liveness, which we could not disable. This is
why Tencent's public other class in Figure~\ref{fig:overview} stays small and
flat. Left in, the health-check stream would account for \num{60}\% of Tencent’s
total packet volume and drown out the \textit{public other} class in
Fig.~\ref{fig:overview}, so we exclude it. As noted in
\S\ref{sec:experience}, our firewall misconfiguration blocked external ICMP on
Tencent, whereas DigitalOcean captured this background Internet traffic. The
missing ICMP is a dataset limitation, as it makes the other class incomparable
across providers.

\paragraph*{Discussion} These results extend CLOUD-NT's provider-dependence
finding eastward and give a first look inside IBR as seen from within Chinese
networks. We observe the same provider-dependent traffic structure in the
Asia-Pacific, and on top of it we see effects a Western-only deployment would
miss, from the campaign aimed at mainland China to the platform's own injected
traffic. 
This is only a \num{20} day two-provider snapshot which can serve as a starting
point for further work.
Because our Tencent nodes sit in China and Southeast Asia and our DigitalOcean
nodes mostly in the West, provider and region largely coincide, and only a
deployment with several providers per region can say how much of what we see is
truly regional.
A community telescope, which we argue for in \S\ref{sec:lessons}, would also
show whether Tencent generalizes to other Chinese providers such as Alibaba, and
whether Western providers in the region (like our DigitalOcean nodes in
Bangalore and Singapore) observe similar traffic patterns.
It provides distributed observation an RI needs to enable transparency in cloud
infrastructures~\cite{hesselmanResponsibleInternetIncrease2020}. Our experience
implies that this structure is needed but hard to build for non-technical
reasons (\S\ref{sec:lessons}).
One finding already stands on its own. The blocklists fit our two providers very
differently. This again hints at targeted scanning and blocklist biases. A
single-provider telescope thus inherits not just a skewed traffic mix but a
blocklist that matches it unevenly.

\section{Lessons and Recommendations}
\label{sec:lessons}

We distill our experience into recommendations for building a diverse
cloud-based network telescope, which we think is needed as an addition not only
to classical network telescopes, but also to the independent observation a
Responsible Internet needs. Our deployment shows that this observation tool is
hard to build. To succeed across a fragmented global network, this observation
infrastructure must be both resilient and responsibly operated.

\paragraph*{For a transparent and resilient Internet}
The Responsible Internet
relies on a Network Inspection Plane to build Internet transparency. This plane
requires independent observers to document Internet security incidents and abuse
\cite{hesselmanResponsibleInternetIncrease2020}. Network telescopes provide this
kind of data by capturing IBR. Classical telescopes, however, miss modern cloud-specific IBR, and cloud-based telescopes fill this gap. The RI explicitly
requires data from multiple vantage points and our findings reinforce this
requirement by showing that IBR depends on the provider and the geo-political
region. Single-region deployments miss events aimed at other regions entirely
and introduce a severe bias. Representative monitoring, therefore, has to be
multi-provider and multi-region. This diversity also makes the observation
infrastructure itself resilient. A network telescope as diverse as the IBR it
captures is harder to evade and survives the loss of any single node. But
building this required diversity proves to be difficult. At 2026 list prices our
whole \num{28}-VM deployment would only cost about \$120, but deployments must
budget not only for VM-hours but for bureaucracy as well. Strict verification
rules can rule out a vantage point entirely (\textbf{L1}) and operating inside
heavily regulated networks requires managing an unreliable network plane
(\textbf{L3}). Furthermore, porting code across divergent ecosystems is not just
a change of credentials, but requires extensive knowledge about platforms and
ecosystems (\textbf{L2}). These barriers, representing the Internet's fragmentation, leave us with the conclusion that no single research group can
clear these hurdles globally. A community-run cloud telescope therefore seems
like the only viable way to pool the onboarding effort and the resulting
visibility, necessary to provide a global view of cloud IBR.

\paragraph*{For a responsibly operated observer}
Cloud-based network telescopes serve as an addition to classical network
telescopes, as shown in~\cite{kempenMultiCloudViewIBR2026}. Therefore they must
use the same passive methods. While reacting to probes is a valid technique
\cite{hiesgenSpokiUnveilingNew2022a}, it alters scanner behavior and breaks our
comparison to classical telescopes. Staying passive also aligns with the RI
mandate to balance transparency against security risks. We minimize data
collection and apply equal care across all regions. This duty is even more
critical for a community telescope pooling shared data. Responsibility also
demands transparency about the biases a deployment introduces. We must therefore
empirically verify what each destination receives and actively disclose
artifacts like blocked ICMP (\S\ref{sec:experience}) or Tencent's internal
probes (\S\ref{sec:vantage}). Unreported biases invalidate the observation.
Ultimately our experience shows how fragmentation impedes global security
research.

\section{Conclusion}
\label{sec:conclusion}

In this report we extended our multi-cloud network telescope CLOUD-NT into the
Asia-Pacific and mainland China. Setting up our telescope with local providers
was hindered by locally determined business practices and strict onboarding
barriers rather than technical design. Our pivot to Tencent and the unreliable
management plane inside China show how Internet fragmentation restricts who can
observe the network (\S\ref{sec:experience}). Once deployed, our new nodes
captured high-volume regional scanning invisible to our Western comparison
deployment (\S\ref{sec:vantage}).

Our experience directly informs the design of a Resilient and Responsible
Internet. Network resilience requires diverse observation infrastructure
to track localized threats. The Responsible Internet framework also depends on
independent measurement to build a transparent Network Inspection Plane. Our
deployment shows that building this global observation layer is bottlenecked by
geo-politics as well as provider restrictions. Overcoming these borders is too
complex for any single research group. We urge the community to collaboratively
build the distributed and responsibly operated monitoring infrastructure an
RI requires.

\bibliographystyle{ACM-Reference-Format}
\bibliography{lib-short}

\end{document}